\documentclass[aps,prb,floatfix,superscriptaddress,twocolumn,footinbib]{revtex4-1}
\usepackage{amssymb}
\usepackage{amsmath}
\usepackage{amsfonts}
\usepackage{appendix}
\usepackage{bm}
\usepackage{graphicx}
\usepackage{epsfig}
\usepackage{epstopdf}
\usepackage{balance}
\usepackage[dvipsnames]{xcolor}
\usepackage{calc}
\usepackage{natbib}
\usepackage[colorlinks,
            linkcolor=blue,
            anchorcolor=blue,
            citecolor=blue,
            urlcolor=blue]{hyperref}
\usepackage{lipsum}
\usepackage{enumerate}

\begin{document}
\title{Moir\'e flat bands in alternating twisted \textrm{MoTe$_2$} multilayer}
\author{Miao Liang}
\affiliation{School of Physics and Wuhan National High Magnetic Field Center, Huazhong University of Science and Technology, Wuhan 430074,  China}
\author{Shi-Ping Ding}
\affiliation{School of Physics and Wuhan National High Magnetic Field Center,
Huazhong University of Science and Technology, Wuhan 430074,  China}
\author{Ming Wu}
\affiliation{School of Physics and Wuhan National High Magnetic Field Center, Huazhong University of Science and Technology, Wuhan 430074,  China}
\author{Chen Zhao}
\affiliation{School of Physics and Wuhan National High Magnetic Field Center, Huazhong University of Science and Technology, Wuhan 430074,  China}
\author{Jin-Hua Gao}
\email{jinhua@hust.edu.cn}
\affiliation{School of Physics and Wuhan National High Magnetic Field Center,
Huazhong University of Science and Technology, Wuhan 430074,  China}
	
\begin{abstract}
The long-awaited fractional quantum anomalous Hall (FQAH) effect recently has been observed in the twisted $\mathrm{MoTe_2}$ homobilayers, causing a great sensation. Here, we theoretically investigate the moir\'e band structures of a closely related system, the alternating twisted multilayer $\mathrm{MoTe_2}$ (ATML-\textrm{MoTe$_2$}), where the adjacent layers have opposite twist angles. We illustrate that such ATML-\textrm{MoTe$_2$} is a very unique moir\'e system, exhibiting multiple topological flat bands highly controllable by the layer number and twist angle, which is not only an ideal platform to simulate Hubbard model, but also may host FQAH states. Specifically, an N-layer ATML-\textrm{MoTe$_2$}  ($N \geq 3$) always possesses $N-2$ topological flat bands near Fermi energy $E_f$, which has an odd-even dependent decomposition rule to understand the behaviors of the moir\'e flat bands. 
We predict three intriguing examples: (1) The AT3L-\textrm{MoTe$_2$} ($N=3$) has one isolated moir\'e flat band, which corresponds to a triangular lattice Hubbard model, resembling the twisted TMD heterobilayers.  (2) The AT4L-\textrm{MoTe$_2$} ($N=4$) has two topological flat bands that are very similar to the twisted $\mathrm{MoTe_2}$ homobilayers, implying the possible existence of FQAH states. (3) When $N>4$, the giant density of states (DOS) induced by the multiple moir\'e flat bands may induce exotic correlated states. 
	\end{abstract}
	
	\maketitle
	\section{Introduction}

Recently, the fractional quantum anomalous Hall (FQAH) effect has been successfully observed in twisted $\mathrm{MoTe_2}$ homobilayers, which has sparked widespread research interest~\cite{cai_signatures_2023,park_observation_2023,zeng_thermodynamic_2023,kang2024evidence,Xu_PRX031037_2023}.
The emergence of the FQAH states in twisted $\mathrm{MoTe_2}$ homobilayers is attributed to its isolated topological moir\'e flat band~\cite{Topolo_Wu_prl2019,Li_Spontaneous_2021,Reddy_Fractional_2023,wang_fractional_2023,Qiu_Interaction_2023,Goldman_Zero_2023,xu_pnas_maximally2024,Reddy_prb245159_2023,yu_fractional_2024,abouelkomsan_band_prb2024,zeng2023integer}. 
Meanwhile, it is further suggested that the uniformly distributed Berry curvature is crucial for the topological flat bands to realize the FQAH states~\cite{Devakul_magic_2021}. All of these important theoretical and experimental studies reveal the uniqueness of the $\mathrm{MoTe_2}$-based moir\'e structures. 
 
The alternating twisted multilayer moir\'e structure is a natural generalization of the twisted homobilayers, where the twist angles of adjacent layers are $\pm \theta$ with opposite rotation directions. The celebrated example is the alternating twisted few-layer graphene (ATFLG)\cite{Khalaf_2019_prb,Carr9b04979,PhysRevB.104.035139,Shin2021,2021Lattice,park_tunable_2021,0Electric2021,cao_pauli-limit_2021,PhysRevResearch.2.033357,xie2022alternating,ma_doubled_2023,Liang_prb_2022,Ding_Mirror195119_2023,Phong_Band_PRB2021,scammell2022theory}, which also host moir\'e flat bands very similar as the twisted bilayer graphene (TBG)\cite{cao2018b,cao2018Correlated,mac2011,Optical_Moon_prb2013,cheng2021}. Interestingly, it is proved that there is a decomposition rule, by which an ATFLG can be decomposed into several effective TBGs with renormalized parameters~\cite{Khalaf_2019_prb}. Therefore, it can be expected that correlated states in TBG will also exist in ATFLG. Note that superconductivity in ATFLG has been observed in recent experiments~\cite{park_tunable_2021,0Electric2021,park_robust_2022,Zhang_promotion_2022,burg_emergence_2022}.
Inspired by the success of ATFLG, a fascinating question arises: can the alternating twisted multilayer $\mathrm{MoTe_2}$ (ATML-\textrm{MoTe$_2$}) generate richer moir\'e band structures, and thus induce more novel correlation and topological effects, such as FQAH states ?

\begin{figure}[ht]
	\centering
	\includegraphics[width=8cm]{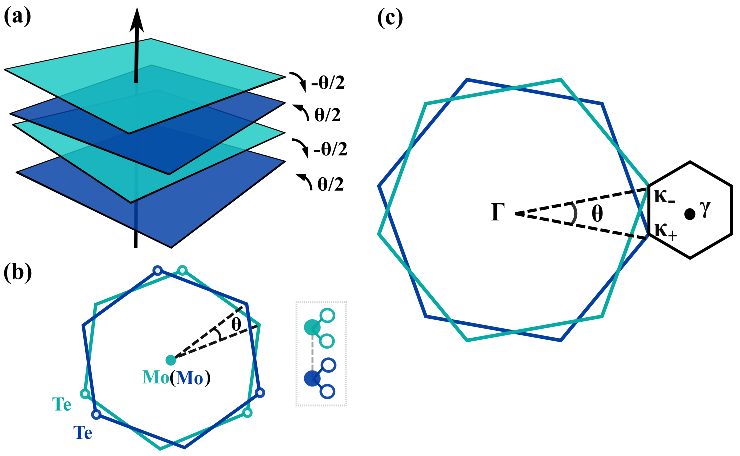}
	\caption{Schematic diagram of ATML-\textrm{MoTe$_2$}. (a) Schematic of AT4L-\textrm{MoTe$_2$}, with the rotation directions of the 1st and 3rd layers (blue) opposite to those of the 2nd and 4th layers (green). (b) Top view of an AA-stacked AT4L-\textrm{MoTe$_2$}. The inset shows a side view between adjacent layers. (c) Brillouin zones of the odd-layer (blue), the even-layer (green), and the moir\'e Brillouin zone (black).
		}\label{stru} 
\end{figure}

In this work, we systematically study the moir\'e band structures of the ATML-\textrm{MoTe$_2$} based on the continuum model method. We find that ATML-\textrm{MoTe$_2$} is a unique moir\'e system hosting multiple topological flat bands, which are highly adjustable by the layer number and twist angle. 
Specifically, an $N$-layer ATML-\textrm{MoTe$_2$} ($N \geq 3$) always has $N-2$ topological flat bands at small twist angle, and the intriguing features of these topological flat bands are summarized as follows: 

\begin{enumerate}
    \item \emph{Multiple degenerate flat bands at small twist angle}. When $\theta \lesssim 1.4^{\circ}$, the $N-2$ flat bands of the $N$-layer ATML-\textrm{MoTe$_2$} are nearly degenerate at $E_f$, with a large band gap separating these flat bands from other low energy bands.
    
      \item \emph{Isolated topological flat band at magic angle}. As $\theta$ increases, the degeneracy of these flat bands is lifted and an isolated flat band appears near the $E_f$. For even $N$, there is a magic angle near $2.02^\circ$, at which an isolated topological flat band is achieved. 

      \item \emph{Decomposition rules}. An even-odd dependent decomposition rule is proved, which can well interpret the origin and behaviors of the $N-2$ topological flat bands.  Meanwhile, there exists a mirror symmetry decomposition for the odd $N$ cases. 
\end{enumerate}

Based on our calculation, we can give three interesting predictions:
 \begin{enumerate}
 \item The AT3L-\textrm{MoTe$_2$} ($N=3$) has one isolated moir\'e flat band near $E_f$, which can be described as a triangular lattice Hubbard model, resembling the twisted transition metal dichalcogenides (TMD) heterobilayers\cite{Hubbard_wu_prl2018,tang2020simulation,Morales_Nonlocal_prl2022}. 

 \item The AT4L-\textrm{MoTe$_2$} ($N=4$) has two topological moir\'e flat bands near $E_f$, which exhibit similar properties to those of the twisted $\mathrm{MoTe_2}$ homobilayers, making it  a promising moir\'e system for achieving the FQAH effect. 

 \item When $N>4$, the presence of multiple degenerate flat bands implies a giant DOS at $E_f$, which is expected to promote the formation of various correlated insulating states. 
 \end{enumerate}

Our results indicate that the ATML-\textrm{MoTe$_2$} is an ideal platform for simulating both the Hubbard model and FQAH states, promising to host a variety of exotic topological and correlated quantum states like that in twisted TMD homobilayers~\cite{cai_signatures_2023,park_observation_2023,zeng_thermodynamic_2023,Xu_PRX031037_2023,Topolo_Wu_prl2019,Li_Spontaneous_2021,Reddy_Fractional_2023,wang_fractional_2023,Qiu_Interaction_2023,Goldman_Zero_2023,xu_pnas_maximally2024,Reddy_prb245159_2023,yu_fractional_2024,abouelkomsan_band_prb2024,zeng2023integer,Devakul_magic_2021,Ab_fang_prb2015,Interlayer_Wang_prb2017,Ultraflat_Naik_prl2018,Zhan_prb241106_2020,wang2020correlated,zhang2020flat,ghiotto2021quantum,Spontaneous_Li_PRR2021,xu2022tunable}. We expect that our predictions can be immediately tested in future experiments. 

The paper is organized as follows. In Sec. II, the model and methods of ATML-\textrm{MoTe$_2$} used in the calculations are given. In Sec. III, we discuss the calculation results. Sec. IV is a short summary. 
 
\section{MODEL AND METHODS}
\subsection{The continuum model of ATML-\textrm{MoTe$_2$}}

In Fig.~\ref{stru}, we give a schematic of ATML-\textrm{MoTe$_2$}. 
The continuum model of ATML-\textrm{MoTe$_2$} can be given in a similar way to that of the twisted TMD homobilayers~\cite{Topolo_Wu_prl2019}. 
For each  $\rm{MoTe_2}$ monolayer, the valence electrons are around the $\pm K$ valleys, which are mainly of the $d$-orbial character with opposite spin due to spin-valley locking\cite{Xiao_prl_Coupled2012,Topolo_Wu_prl2019}. Considering the layer-dependent potential and interlayer tunneling, the Hamiltonian of $N$-layer ATML-\textrm{MoTe$_2$} of the $+K$ valley is 
	\begin{equation}
		\begin{aligned}\label{H_N}
			\mathcal{H} =\left(\begin{array}{ccc}
				\mathcal{H}_1 & T_{b m} & 0 \\
				T_{b m}^{+} & \mathcal{\widetilde{H}}_{m} & T_{t m} \\
				0 & T_{t m}^{+} & \mathcal{H}_N
			\end{array}\right),
		\end{aligned}
	\end{equation}
where $\mathcal{H}_1$ and $\mathcal{H}_N$ describe the bottom (1st) and top (Nth) TMD layer, respectively.  $\widetilde{H}_m$ represents the middle $N-2$ layers.
 
For the outermost two layers, 
	\begin{equation}
		\begin{aligned}\label{H_1}
			\mathcal{H}_{1/N}=-\frac{\hbar^2\left(\mathbf{k}-\mathbf{\kappa}_{1/N} \right)^2}{2 m^*}+\Delta\left(\mathbf{r}, \theta_{1/N}\right),
		\end{aligned}
	\end{equation}
where $\theta_{1/N}$ is the twist angle of the 1st/Nth monolayer, and $\mathbf{\kappa}_{1/N}$ is the corresponding momentum offset. As shown in Fig.~\ref{stru}, for the alternating twist case, we always assume the twist angle of the odd (even) layer is $\theta/2$ ($-\theta/2$), and the momentum offset is $\mathbf{\kappa}_+$ ($\mathbf{\kappa}_-$) for the odd (even) layer, where $\mathbf{\kappa}_{\pm}=\left[4 \pi /\left(3 a_M\right)\right](-\sqrt{3} / 2, \mp 1 / 2)$. $a_M=a_0/\theta$ is the period of the moir\'e potential, and  $a_0$ is the lattice constant of the TMD monolayer.
 $\Delta\left(\mathbf{r}, \theta \right)$  describes the layer-dependent moir\'e potential 
	\begin{equation}
		\begin{aligned}
			\Delta(\mathbf{r}, \theta)=2 V \sum_{j=1,3,5} \cos \left(\mathbf{g}_j \cdot \mathbf{r}+\eta \psi\right)	\end{aligned},
	\end{equation}
where $\mathbf{g}_j$ is $(j-1)\pi/3$ counterclockwise rotation of the moir\'e reciprocal lattice vector $\mathbf{g}_1 = (4\pi /\sqrt{3} a_M,0)$.   $\eta(\theta)=\mathrm{sign}(\theta)$ denotes the direction of the rotation, e.g.~$\eta(\theta/2)=1$. For $\rm{MoTe_2}$, we set $a_0=3.472 \mathring{\rm{A}}$, $V=8~\rm{meV}$, $\psi=-89.6^{\circ}$, and the effective mass $m^* =0.62~m_e$~\cite{Topolo_Wu_prl2019,Three_prb085433_2013}. 

For the middle $N-2$ layers, the Hamiltonian  $\mathcal{\widetilde{H}}_m$ is,
	\begin{equation}
		\begin{aligned}\label{H_m}
			\mathcal{\widetilde{H}}_{m}=\left(\begin{array}{ccccc}
				\mathcal{H}_2 & \Delta_T^{\dagger} (\mathbf{r})& 0 & \cdots & 0 \\
				\Delta_T(\mathbf{r}) & \mathcal{H}_3 & \Delta_T(\mathbf{r}) & \cdots & 0 \\
				0 & \Delta_T^{\dagger}(\mathbf{r}) & \mathcal{H}_4 & \cdots & 0 \\
				\vdots & \vdots & \vdots & \ddots & \vdots \\
				0 & 0 & 0 & \cdots & \mathcal{H}_{N-1}
			\end{array}\right)
		\end{aligned}
	\end{equation}
 where $\mathcal{H}_i$ is the Hamiltonian of the i-th layer with $1<i<N$.
	\begin{equation}\label{hmiddlelayer}
		\begin{aligned}
			\mathcal{H}_i=-\frac{\hbar^2\left(\mathbf{k}-\mathbf{\kappa}_{i}\right)^2}{2 m^*}+2 \Delta\left(\mathbf{r}, \theta_i\right)
		\end{aligned}
	\end{equation}
where $i= 2, 3,...,N-1$ is the layer index.  $\mathbf{\kappa}_i$ and $\theta_i$ are the momentum offset and twist angle of the ith layer as defined before. Note that the $i$th layer here feels the moir\'e potential from both the $(i-1)$th and $(i+1)$th layers, which thus is distinct from the top and bottom layers in Eq.~\eqref{H_1}. 
$\Delta_{\mathrm{T}}$ is the interlayer tunneling  
	\begin{equation}
		\begin{aligned}
			\Delta_T(\mathbf{r})=w\left(1+e^{-i \cdot \mathbf{g}_2 \cdot \mathbf{r}}+e^{-i \cdot \mathbf{g}_3 \cdot \mathbf{r}}\right). 
		\end{aligned}
	\end{equation}
where $w=-8.5$ meV is the interlayer tunneling strength~\cite{Topolo_Wu_prl2019}.
 
The interlayer tunneling terms in Eq.~\eqref{H_N} are
	\begin{equation}
		\begin{aligned}
			&T_{b m}=\left(\Delta_T(\mathbf{r}), 0, \ldots, 0\right) \\
			&T_{t m}=\left(0, \ldots, 0,\Delta_T(\mathbf{r})\right)
		\end{aligned}
	\end{equation}

\subsection{Decomposition rules}
Firstly, for an $N$-layer ATML-\textrm{MoTe$_2$}, it should be noted that its electronic states near $E_f$, i.e., the moir\'e flat bands, mainly originate from the middle $N-2$ layers $\widetilde{H}_m$, with little influence from the outermost two layers. It is because, as shown in Eq.~\eqref{H_1} and \eqref{hmiddlelayer}, the moir\'e potential applied on the outermost two layers is $\Delta$, while that of the middle layers is $2\Delta$.  
The difference of moir\'e potential is significant, causing the electronic states of the outermost two layers to be far from the $E_f$. Therefore, it is a good approximation to use the electronic structure of $\widetilde{H}_m$ to understand the moir\'e band structures of ATML-\textrm{MoTe$_2$} near $E_f$, where the influence of the outermost two layers can be disregarded. The numerical results show that such an approximation works quite well when $\theta$ is small. Hereafter, we always use $N$ to refer to the total number of layers in an ATML-\textrm{MoTe$_2$}.

Very interestingly, $\widetilde{H}_m$ here has an even-odd dependent decomposition rule, due to its alternating twist configuration. The decomposition rule here is very similar to that of the ATFLG. First, the middle $N-2$ layers in $\widetilde{H}_m$ can be divided into two sets, i.e.~$\{H_i | i\in even\}$ and $\{H_i | i\in odd\}$, where $H_i$ are identical in each set due to the alternating twist structure. As is proved in Ref.~\onlinecite{ledwith_tb_2021}, such kind of Hamiltonian matrix can be decomposed using singular value decomposition scheme. This decomposition depends on whether the layer number of $\widetilde{H}_m$, i.e.~$(N-2)$, is even or odd: 
\begin{equation}\label{decouple}
    \mathcal{\widetilde{H}}^{dec}_{m}= 
        \begin{cases}
            \oplus^n_{\nu=1}h(\lambda_\nu), & \text{if N-2 is even},           \\    \oplus^n_{\nu=1}h(\lambda_\nu)\oplus\mathcal{H}_{2}, &  \text{if N-2 is odd}, 
        \end{cases}
\end{equation}
where $\lambda_\nu=2 \cos \frac{\pi \nu}{N-1}$ with $\nu=1,...,n$ and $n=[(N-2)/2]$. Here,  $h(\lambda_\nu)$ is an effective twisted TMD homobialyer with coupling parameter $\lambda_\nu$
\begin{equation}
    \begin{aligned}\label{h}
	h(\lambda_\nu)=\left(\begin{array}{cc}
	\mathcal{H}_2 & \lambda_{\nu}\Delta_T(\mathbf{r})^{\dagger} \\
	\lambda_{\nu} \Delta_T(\mathbf{r}) & \mathcal{H}_{3} 
		\end{array}\right).
	\end{aligned}
\end{equation}
where $H_{2}$ actually represents a twisted heterobilayers, i.e.~an effective TMD monolayer under a moir\'e potential as given in Eq.~\eqref{hmiddlelayer}.  
Note that, in Eq.~\eqref{decouple} and \eqref{h}, $H_2$ ($H_3$) is used to denote an effective TMD monolayer with a twist angle $-\theta/2$ ($\theta/2$).
The decomposition above implies that, near $E_f$,  the moir\'e bands of $N$-layer ATML-\textrm{MoTe$_2$} can be approximately decoupled into those of several twisted TMD homobilayers (and a  TMD monolayer with a moir\'e potential) if $N$ is even (odd). 


Furthermore,  each $h(\lambda_\nu)$, as a twisted TMD homobilayers, will provide two moir\'e flat bands near $E_f$, and the $H_2$ in Eq.~\eqref{decouple} will offer one isolated moir\'e flat band at $E_f$, similar as the twisted TMD heterobilayers. So, the decomposition rules in Eq.~\eqref{decouple} actually imply that, with a small $\theta$, an N-layer ATML-\textrm{MoTe$_2$} should always exhibit $N-2$ moir\'e flat bands near $E_f$. 

In addition to the decomposition rule with Eq.~\eqref{decouple}, an $N$-layer ATML-\textrm{MoTe$_2$} with odd $N$ exhibits mirror symmetry relative to the middle monolayer, resulting in a mirror symmetry decomposition as well\cite{Ding_Mirror195119_2023,zhang2023chiral}.  It should be noted that the mirror symmetry decomposition here is exact, but the decomposition in Eq.~\eqref{decouple} is only valid for the moir\'e bands very near $E_f$.

\section{RESULTS AND DISCUSSION}	

In this section, we show our numerical results about the ATML-\textrm{MoTe$_2$}, where the cases of even and odd $N$ are discussed separately. 

\subsection{Odd-layer ATML-\textrm{MoTe$_2$}}

\subsubsection{AT3L-\textrm{MoTe$_2$}}

We first consider the case of AT3L-\textrm{MoTe$_2$} ($N=3$). 
As we will see, an AT3L-\textrm{MoTe$_2$} always has one moir\'e flat band near $E_f$, which is approximately equivalent to a triangle lattice Hubbard model resembling the twisted TMD heterobilayers when $\theta$ is small. 

The AT3L-\textrm{MoTe$_2$} has a mirror symmetry relative to the middle monolayer. So, as mentioned in the last section, we can do a mirror symmetry decomposition for its moir\'e bands.  
To do the mirror symmetry decomposition, we first have to build a parity-resolved basis $\{|\phi_{1,3},e\rangle,|\phi_2,e\rangle,|\phi_{1,3},o\rangle\}$ according to the mirror symmetry, where
\begin{equation}
		\begin{aligned}
			&|\phi_{1,3},e\rangle=\frac{1}{\sqrt{2}}(|\phi_1\rangle+|\phi_3\rangle),\\	
    &|\phi_{2},e\rangle=|\phi_{2}\rangle,\\
   &|\phi_{1,3},o\rangle=\frac{1}{\sqrt{2}}(|\phi_1\rangle-|\phi_3\rangle).
		\end{aligned}
	\end{equation} 
Here, $e/o$ denotes the even/odd parity relative to the middle monolayer.  $|\phi_{1,2,3}\rangle$ are the original basis functions of the 1st, 2nd, 3rd monolayers, used in the Hamiltonian~\eqref{H_N}. 
 This parity-resolved basis gives rise to a transformation matrix,
	\begin{equation}
		\begin{aligned}
			U=\frac{1}{\sqrt{2}}\left(\begin{array}{ccc}
				1 & 0 & 1 \\
				0 & \sqrt{2} & 0 \\
				1 & 0 & -1
			\end{array}\right).
		\end{aligned}
	\end{equation}
Under this transformation, we get a decoupled Hamiltonian $\mathcal{H}_{mirr}^{\prime}=U^{-1}\mathcal{H}U$, where 
	\begin{equation}\label{Hmirr}
		\begin{aligned}
			\mathcal{H}_{mirr}^{\prime}=\left(\begin{array}{ccc}
				\mathcal{H}_1 & \sqrt{2}\Delta_T(\mathbf{r}) & 0 \\
				\sqrt{2}\Delta_T^{\dagger}(\mathbf{r}) &  \mathcal{H}_2 & 0 \\
				0 & 0 & \mathcal{H}_3
			\end{array}\right).
		\end{aligned}
	\end{equation}


\begin{figure*}[ht]
		\centering
		\includegraphics[width=17cm]{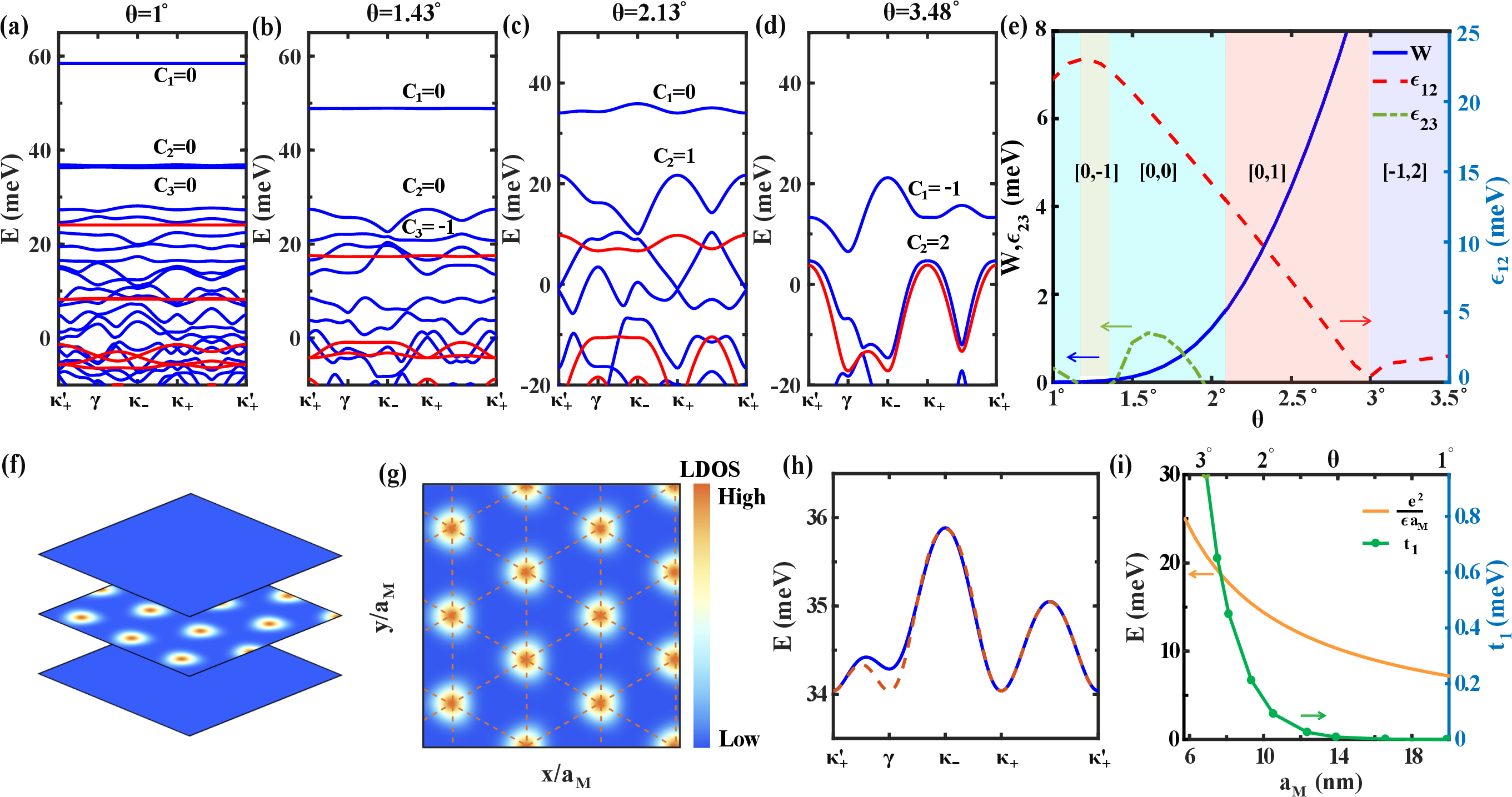}
		\caption{(a)-(d) The moir\'e band structures of AT3L-\textrm{MoTe$_2$} for (a) $\theta = 1^\circ$, (b) $1.43^\circ$, (c) $2.13^\circ$, and (d) $3^\circ$. Blue (red) lines represent the bands with even (odd) parity. 
(e) The bandwidth W for the 1st band, and the energy gap $\epsilon_{12}$ ($\epsilon_{23}$) between the 1st (2nd) and 2nd (3rd) bands as a function of $\theta$. By using different colors to identify the various regions of $\theta$, the top two moir\'e bands have distinct Chern numbers $[C_1,C_2]$ within these regions. (f) shows the 1st flat band wave function distribution in each layer of the AT3L-\textrm{MoTe$_2$}, which correspond to the $\gamma$ point for $\theta=2.13^\circ$. (g) We present the wave function distribution of the middle layer. (h) The blue solid line is the 1st band of AT3L-\textrm{MoTe$_2$} at $\theta=2.13^\circ$, which is the enlarged view of the 1st band in (c). The orange dashed line is a tight-binding-model fit to this band that includes hopping up to the 3rd nearest-neighbor. (i) The yellow solid line is a characteristic interaction energy scale for $\epsilon=10$, and the green solid line is the nearest-neighbor hopping parameter $t_1$ as a function of moir\'e period $a_M$.
 }\label{3a} 
\end{figure*}




We see that the Hamiltonian of AT3L-\textrm{MoTe$_2$} is now decoupled into two subsystems with opposite parity. Formally, the part with even parity is very like twisted TMD homobilayers with the interlayer tunneling scaled by $\sqrt{2}$, while the part with odd parity is equivalent to a TMD monolayer with an applied moir\'e potential, i.e.~$H_3$. An important fact is that the moir\'e potentials in $H_{1,3}$ are largely different from that in $H_2$, so that the electron states from $H_{1,3}$ are pushed far away from the $E_f$, especially when $\theta$ is small. Therefore, at a small twist angle, the moir\'e bands of AT3L-\textrm{MoTe$_2$} near $E_f$ mainly originate the middle monolayer $H_2$ with even parity. If we further note that $H_2$ here is nearly identical to the Hamiltonian of a twisted TMD heterobilayers (e.g.~$\mathrm{WSe_2/WS_2}$) in form, which hosts one isolated moir\'e flat band at $E_f$, described by a triangle lattice Hubbard model\cite{Hubbard_wu_prl2018,tang2020simulation,Morales_Nonlocal_prl2022}. It can be expected that, when $\theta$ is small, an AT3L-\textrm{MoTe$_2$} will have a single isolated moir\'e flat band (even parity) near $E_f$, similar to the one in $\mathrm{WSe_2/WS_2}$ system, which can be described by a triangle lattice Hubbard model. 

However, we emphasize that this picture of the triangle lattice Hubbard model is only valid for small twist angle regions. It is because, as $\theta$ approaches $3^\circ$, the moir\'e band from $H_{1,3}$ in Eq.~\eqref{Hmirr} will approach that of $H_2$, and the hybridization between them cannot be ignored. In other words,  when $\theta$ becomes large enough, the part with even parity in Eq.~\eqref{Hmirr} will behave more like a twisted TMD homobilayers, instead of a twisted TMD heterobilayers, and the 1st moir\'e band near $E_f$ will get a nonzero Chern number then.  

The numerical results of the moir\'e bands for AT3L-\textrm{MoTe$_2$} with various twist angles are given in Fig.~\ref{3a}, where blue (red) lines represent the bands with even (odd) parity.
Fig.~\ref{3a} (a) shows the moir\'e band structure at  $\theta=1^\circ$. As we expected, at such small $\theta$, we do get an isolated moir\'e flat band near $E_f$, which is separated from other moir\'e bands by a large gap of about $20~\rm meV$. Meanwhile, the odd parity bands (red lines) are far away from the $E_f$, so that the three topmost moir\'e bands are all with even parity (blue lines). According to the analysis above, we in fact can consider that the three topmost moir\'e bands mainly originate from an equivalent twisted TMD heterobilayers like 
$\mathrm{WSe_2/WS_2}$, where the first one is a moir\'e $s$-band.
The numerical results indicate that these three moir\'e bands are topologically trivial, i.e.~Chern number is zero, when $\theta$ is small.

The bandwidth of the 1st moir\'e band $W$ increases monotonically as $\theta$ increases, see Fig.~\ref{3a} (e). We see that when $\theta<1.5^\circ$, the 1st moir\'e band is almost completely flat as shown in Fig.~\ref{3a} (b). In Fig.~\ref{3a} (e), we also plot the gap between the 1st moir\'e band and other low energy bands, i.e. $\epsilon_{12}$, as a function of $\theta$. And the gap between the 2nd and 3rd moir\'e bands $\epsilon_{23}$ is also shown in Fig.~\ref{3a} (e) as well.  
As $\theta$ becomes larger, $\epsilon_{12}$ nearly monotonically decreases, and it becomes zero when $\theta$ approaches $3^\circ$. Thus, in the regime of $\theta<3^\circ$, the 1st moir\'e band remains isolated and its Chern number stays at zero, see Fig.~\ref{3a} (c). 
 
Due to the similarity with the $\mathrm{WSe_2/WS_2}$ system, we expect that the 1st moir\'e band here should be approximately equivalent to a triangle lattice Hubbard model when $\theta$ is small. To verify this prediction, we first calculate the wave function of the 1st moir\'e flat band ($\gamma$ point, $\theta=2.13^\circ$), as shown in Fig.~\ref{3a} (f) and (g). The numerical results show that the wave function distribution of the 1st moir\'e band is mainly localized in the middle monolayer, and the wave function distribution of clearly forms a triangle lattice, as we expect. In this situation, the influence of the outermost two layers, i.e.~$H_{1,3}$ in Eq.\eqref{Hmirr}, is small, so that the triangle lattice Hubbard model is a rather good approximation. This is also demonstrated clearly by the fitting of the 1st moir\'e band with a tight binding model of triangle lattice, as given in Fig.~\ref{3a} (h). The characteristic Coulomb energy $\frac{e^2}{\epsilon a_M}$, which is proportional to the on-site Hubbard $U_0$, is plotted in Fig.~\ref{3a} (i). And, the nearest-neighbor (NN) hopping parameter $t_1$ of the triangle lattice model got by fitting the 1st moir\'e band as a function of moir\'e period $a_M$ is given in Fig.~\ref{3a} (i) as well. Note that the exotic correlated states in twisted TMD heterobilayers have been intensively studied recently\cite{Hubbard_wu_prl2018,seyler2019signatures,jin2019observation,tran2019evidence,Wufengcheng2018theory,wu2017topological,xu2020correlated,Slagle_Charge_Prb2020,tang2020simulation,huang2021correlated,li2021quantum,regan2020mott,kennes2021moire,li2021continuous,Morales_Nonlocal_prl2022,wang2022light,xie2023nematic}.

With a larger $\theta$, the 2nd and 3rd moir\'e bands can have a nonzero Chern number once touching with other moir\'e bands occurs. Note that the Chern number hereafter is for one valley, and the other valley has the opposite Chern number due to the time reversal symmetry. 
In Fig.~\ref{3a} (b), the Chern number of the 3rd moir\'e band becomes $C_3=-1$ when $\theta=1.43^\circ$. And with $\theta=2.13^\circ$, the 2nd moir\'e band has a nonzero Chern number $C_2=1$, see Fig.~\ref{3a} (c). Interestingly, as $\theta$ increases and crosses $3^\circ$, the gap $\epsilon_{12}$ will first close and reopen, so that the 1st moir\'e band gets a nonzero Chern number $C_1=-1$ once $\theta>3^\circ$, see Fig.~\ref{3a} (d). As analyzed above, in this case, the triangle lattice Hubbard model becomes invalid for the 1st moir\'e band.  
Based on our numerical calculations above, we then can identify various regions of $\theta$ in Fig.~\ref{3a} (e), denoted by different colors, in which the moir\'e bands have distinct Chern numbers $[C_1,C_2]$ (the 3rd moir\'e band is hard to be accessed in the experiment). 

Finally, we would like to mention that the moir\'e band structures of  AT3L-\textrm{MoTe$_2$} have been also discussed in two other recent papers, but the numerical results are analyzed from a completely different perspective~\cite{li2023tuning,AlBuhairan_prb2023}.  The decomposition of moir\'e bands, the analysis of the origination of the 1st moir\'e band, and the prediction about the equivalent Hubbard model are the focus of this work. 
Meanwhile, we also notice the twisted TMD homotrilayer systems have been successfully fabricated in recent experiments~\cite{zheng2023exploring,liu2023inversion,leng2021intrinsic,liao2020precise,zhong2024trilayer}. 

\begin{figure*}[ht]
		\centering
		\includegraphics[width=17cm]{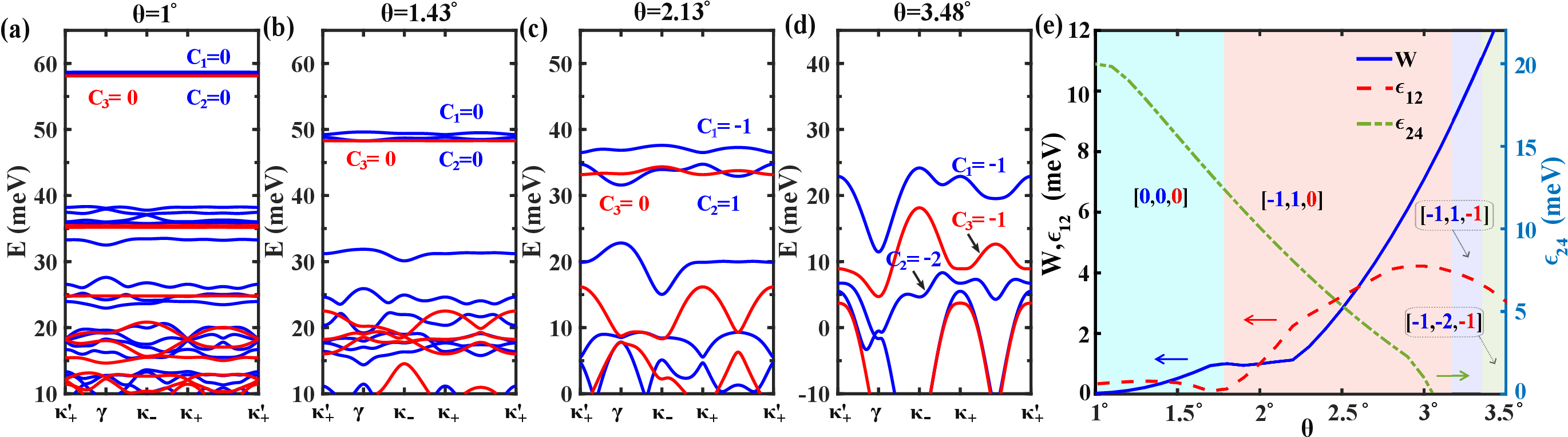}
		\caption{The moir\'e band structures of AT5L-\textrm{MoTe$_2$} for (a) $\theta = 1^\circ$, (b) $1.43^\circ$, (c) $2.13^\circ$, and (d) $3.48^\circ$. Blue (red) lines represent the bands with even (odd) parity. 
In (e), we plot $W$, $\epsilon_{12}$ and $\epsilon_{24}$ as functions of $\theta$. $W$ is the bandwidth of the 1st band. $\epsilon_{12}$ ($\epsilon_{24}$) is the gap between the 1st (2nd) and 2nd (4th) bands. Here, the 1st, 2nd and 4th bands refer to the three topmost even parity bands (blue lines), while the 3rd band is the topmost odd parity band (red lines). The regions with different Chern numbers $[C_1,C_2,C_3]$ are denoted by different colors. 
      }\label{5a} 
\end{figure*}

\subsubsection{AT5L-\textrm{MoTe$_2$}}
The case of AT5L-\textrm{MoTe$_2$} ($N=5$) is more complex. According to the decomposition rule in Eq.~\eqref{decouple}, we can expect that, when $\theta$ is not too large,  AT5L-\textrm{MoTe$_2$} should have three  moir\'e flat bands near $E_f$ isolated from other low energy bands, which are mainly from the middle three monolayers.

Note that all the ATML-\textrm{MoTe$_2$} with odd N have mirror symmetry relative to the middle monolayer. So, we can also do the mirror symmetry decomposition for the AT5L-\textrm{MoTe$_2$}. Similarly, we first needed to build a parity-resolved basis $\{|\phi_{1,5},e\rangle,|\phi_{2,4},e\rangle,|\phi_{3},e\rangle,|\phi_{2,4},o\rangle|\phi_{1,5},o\rangle\}$, where
\begin{equation}
		\begin{aligned}
			&|\phi_{1,5},e\rangle=\frac{1}{\sqrt{2}}(|\phi_1\rangle+|\phi_5\rangle)\\	
          &|\phi_{3},e\rangle=|\phi_3\rangle\\	
            &|\phi_{1,5},o\rangle=\frac{1}{\sqrt{2}}(|\phi_1\rangle-|\phi_5\rangle)
		\end{aligned}
	\end{equation}
Based on this basis, we get the following transformation matrix,
	\begin{equation}
		\begin{aligned}
			U=\frac{1}{\sqrt{2}} \left(\begin{array}{ccccc}
				1 & 0 & 0 & 0 & 1 \\
				0 & 1 & 0 & 1 & 0 \\
				0 & 0 & \sqrt{2} & 0 & 0 \\
				0 & 1 & 0 & -1 & 0 \\
				1 & 0 & 0 & 0 & -1 \end{array}\right).
		\end{aligned}
	\end{equation}	
After the transformation, a decoupled Hamiltonian is achieved
	\begin{equation}
		\begin{aligned}
        &\mathcal{H}_{mirr}^{\prime}\\
              =&\left(\begin{array}{ccccc}
				\mathcal{H}_1 & \Delta_T(\mathbf{r}) & 0 & 0 & 0\\
				\Delta_T^{\dagger}(\mathbf{r}) & \mathcal{H}_2 & \sqrt{2}\Delta_T^{\dagger}(\mathbf{r}) & 0 & 0\\
				0 & \sqrt{2} \Delta_T(\mathbf{r}) & \mathcal{H}_3 & 0 & 0 \\
				0 & 0 & 0 & \mathcal{H}_4 & \Delta_T^{\dagger}(\mathbf{r}) \\
				0 & 0 & 0 & \Delta_T (\mathbf{r})& \mathcal{H}_5
			\end{array}\right).
		\end{aligned}
	\end{equation}
In this decoupled Hamiltonian, the part with even parity is equivalent to an alternating twisted TMD trilayer with asymmetric interlayer tunneling, where the one between $H_2$ and $H_3$ is scaled by $\sqrt{2}$. Meanwhile, the odd parity part is an equivalent twisted TMD bilayer. 
 
Similar to that in AT3L-\textrm{MoTe$_2$}, the moir\'e potentials in the outermost two layers ($H_{1,5}$) are significantly different from those in the middle three monolayers, which makes the electronic states from $H_1$ and $H_5$ far from the $E_f$, especially at small twist angle. So, in the small twist angle region, the even parity part will behave like the twisted TMD homobilayers, which offers a pair of moir\'e flat bands near $E_f$. Similarly, the odd parity part will host one isolated moir\'e flat band at $E_f$, which originates from the $H_4$. Therefore, the mirror symmetry decomposition here indicates that AT5L-\textrm{MoTe$_2$} exhibits three moir\'e flat bands at $E_f$ for a small twist angle, which is in good agreement with the deduction of the decomposition rule in Eq.~\eqref{decouple}. Actually, Eq.~\eqref{decouple} also indicates that AT5L-\textrm{MoTe$_2$} can be decoupled into one twisted TMD homobilayers $h(\lambda_{\nu=1})$ and one twisted heterobilayers described as $H_4$. 

The numerical results of the moir\'e bands of  AT5L-\textrm{MoTe$_2$}  are given in Fig.~\ref{5a}. In Fig.~\ref{5a} (a-d), we plot the moir\'e bands of AT5L-\textrm{MoTe$_2$} for different twist angles, where the blue (red) lines are for the even (odd) parity. As expected, when $\theta$ is small, AT5L-\textrm{MoTe$_2$} does have three moir\'e flat bands at $E_f$, isolated from other low energy bands by an obvious gap, see Fig.~\ref{5a} (a-c). Here, two of the three moir\'e flat bands with even parity (blue lines) behave like that of a twisted \textrm{MoTe$_2$} homobilayers, while the left one (red lines, odd parity) corresponds to the moir\'e flat band of a twisted TMD heterobilayers.

At a small twist angle $\theta=1^\circ$, the three moir\'e flat bands are nearly degenerate at $E_f$, see Fig.~\ref{5a} (a). Such threefold degeneracy of the moir\'e flat bands here will give rise to a giant DOS at $E_f$, which may induce exotic correlated insulating states in experiments.  
Increasing $\theta$, the bandwidth of the moir\'e flat bands increases and a gap between the two even parity flat bands (blue lines) becomes observable, as shown in Fig.~\ref{5a} (b) and (c). In the region  $3.2^\circ > \theta>1.8^\circ$ [Fig.~\ref{5a} (b) and (c)], the Chern number of the two even parity flat bands becomes nonzero, i.e.~$[C_1,C_2]=[-1,1]$ like that in twisted \textrm{MoTe$_2$} homobilayers, while that of the odd parity flat band (red lines) remains zero, i.e.~$C_3=0$. When  $\theta$ is large enough, e.g.~$\theta=3.48^\circ$ in Fig.~\ref{5a} (d), the three moir\'e flat bands are no longer isolated from other low energy bands, and the Chern numbers are changed accordingly, i.e.~$[C_1,C_2,C_3]=[-1,-2,-1]$. 
 
In Fig.~\ref{5a} (e), we plot the $W$, $\epsilon_{12}$ and $\epsilon_{24}$ as  functions of $\theta$. Here, $W$ is the bandwidth of the 1st moir\'e flat band, $\epsilon_{12}$ is the gap between the two even parity moir\'e flat bands, and $\epsilon_{24}$ represents the gap between the 2nd moir\'e flat band and the 4th moir\'e band (blue, even parity). We also denote the regions of $\theta$ with different Chern numbers by different colors, as shown in Fig.~\ref{5a} (e). We see that, though $W$ increases monotonically with $\theta$, the AT5L-\textrm{MoTe$_2$} does not have a clear magic angle. In principle, in the region $1.8^\circ<\theta<2.2^\circ$, we can get an isolated moir\'e flat band at $E_f$, separated from other moir\'e bands with a small but detectable gap. 

\subsection{Even-layer ATML-\textrm{MoTe$_2$}}

In this section, we then discuss the moir\'e band structures of the ATML-\textrm{MoTe$_2$} with even $N$. 

\begin{figure*}[ht]
		\centering
		\includegraphics[width=17cm]{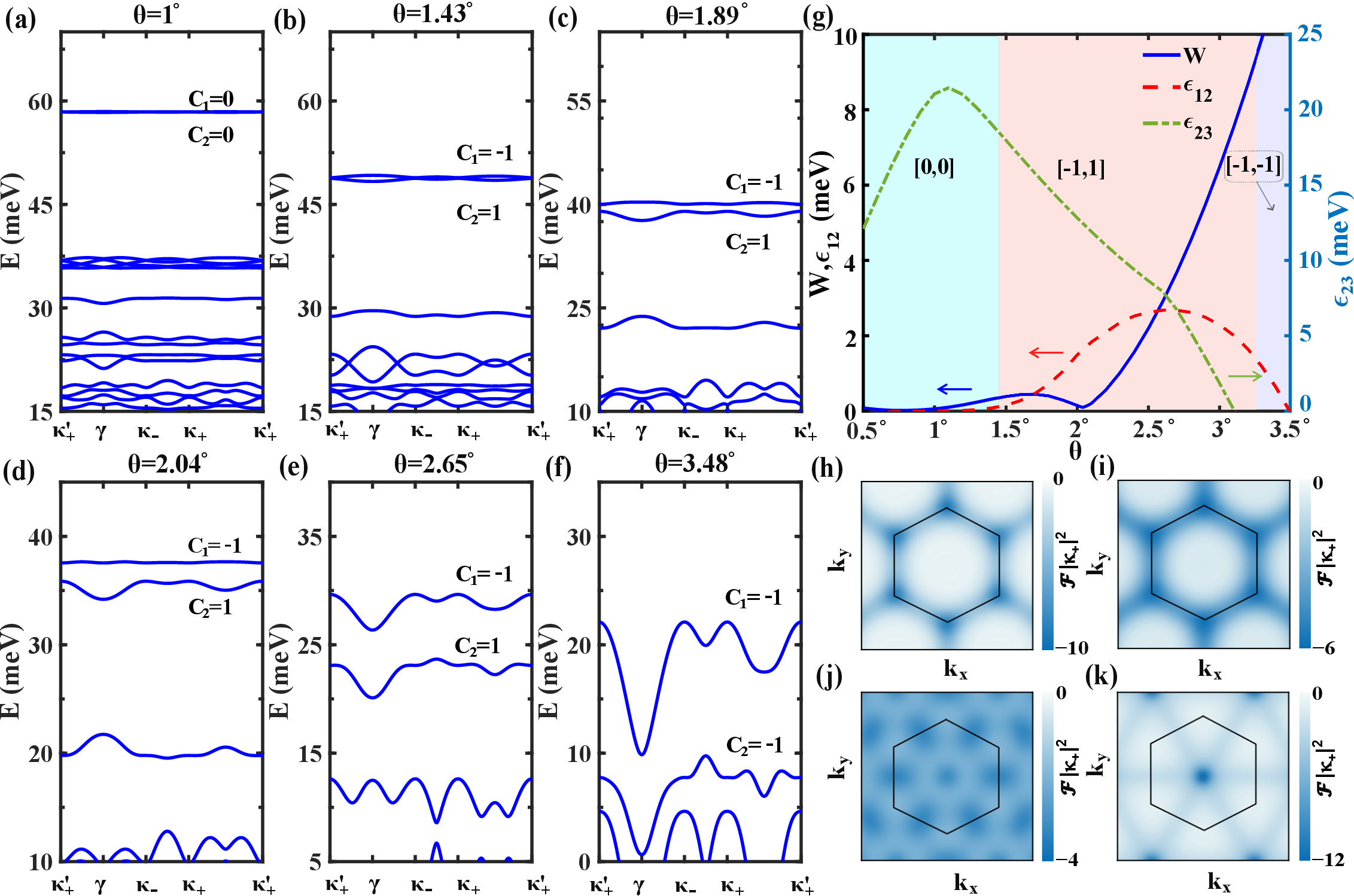}
		\caption{(a)-(f) The moir\'e band structures of AT4L-\textrm{MoTe$_2$} for twist angle $\theta =1^{\circ},1.43^{\circ},1.89^{\circ},2.04^{\circ},2.65^{\circ},3^{\circ}$. (g) The bandwidth W of the 1st band(solid blue line), the gap $\epsilon_{12}$ between the 1st and 2nd bands (dashed red line), and the gap $\epsilon_{23}$ between the 2nd and 3rd bands (dashed green line) as a function of twist angle. Different colored regions represent different Chern numbers of the top two bands. (h)-(k) show the distribution of Berry curvature $\mathcal{F}$ for (h) $\theta =1.89^{\circ}$, (i) $\theta =2.04^{\circ}$, (j) $\theta =2.65^{\circ}$, and (k) $\theta =3.48^{\circ}$.
     }\label{4a} 
\end{figure*}

\subsubsection{AT4L-\textrm{MoTe$_2$}}

We first consider the case of AT4L-\textrm{MoTe$_2$}. The numerical results are shown in Fig.~\ref{4a}, where the moir\'e band structures of AT4L-\textrm{MoTe$_2$} at different twist angles are given, as illustrated in Fig.~\ref{4a} (a)-(f). 

For AT4L-\textrm{MoTe$_2$}, the most prominent feature is that, at a small twist angle, it always exhibits two moir\'e flat bands at $E_f$, which are separated from the remaining moir\'e bands by a large gap. This is quite like the twisted $\mathrm{MoTe_2}$ homobilayers, as indicated by the decomposition rules in Eq.~\eqref{decouple}. Note that the ATML-\textrm{MoTe$_2$} with even $N$ does not have mirror symmetry anymore. 
The two moir\'e flat bands exhibit distinct behaviors across different regions of $\theta$, and three regions of $\theta$ are identified according to their distinct topological features: 

When $\theta \lesssim 1.4^{\circ}$, the two moir\'e flat bands are nearly degenerate at the $E_f$, which are separated from other bands by a large gap. The Chern numbers of the first two flat bands are $[C_1,C_2]=[0,0]$ in this region, as indicated in Fig.~\ref{4a} (a) and (g). The case of $\theta=1^\circ$ is plotted in Fig.~\ref{4a} (a). As shown in Fig.\ref{4a}(g), the minimum value of the bandwidth $W$ of the first flat band occurs around $\theta=1^\circ$. Meanwhile, $\epsilon_{23}$ has its maximum value about  $21.48~\rm{meV}$ around $\theta \approx 1.1^\circ$. 
 
Increasing $\theta$ to the region $1.4^\circ < \theta <3.3^\circ$, the degeneracy of the two moir\'e flat bands is lifted, and the gap $\epsilon_{12}$ becomes obvious. Meanwhile, the Chern numbers of the two moir\'e bands become nonzero, i.e.~$[-1,1]$. We thus can get an isolated topological flat band near $E_f$. In this region, we plot $\theta=1.43^\circ$ in Fig.~\ref{4a}(b), where a sudden change of the Chern number occurs near $\theta \approx 1.4^\circ$. The gap between the two moir\'e flat bands is observable when $\theta=1.89^\circ$, see Fig.~\ref{4a}(c). Most interestingly, we find a magic angle around $\theta \approx 2.02^\circ$, where the bandwidth of the 1st moir\'e flat band becomes as small as $0.1$ meV, as shown in Fig.~\ref{4a}(d) and (g). Further increasing $\theta$ to $2.65^\circ$, the two moir\'e bands become more dispersive, see Fig.~\ref{4a}(e). When $\theta$ approaches about $3.1^\circ$, the gap $\epsilon_{23}$  becomes zero, see Fig.~\ref{4a} (g).

In the region of $\theta>3.3^\circ$, we find a change of Chern number from $[-1,1]$ to $[-1,-1]$, see Fig.~\ref{4a}(f) and (g). In this region, the bandwidth of the first two moir\'e bands will increase monotonically as $\theta$ increases. Consequently, the two bands gradually begin to overlap in energy, i.e.~$\epsilon_{12}=0$, though they are still isolated from each other. 

\begin{figure*}[ht]
		\centering
		\includegraphics[width=17cm]{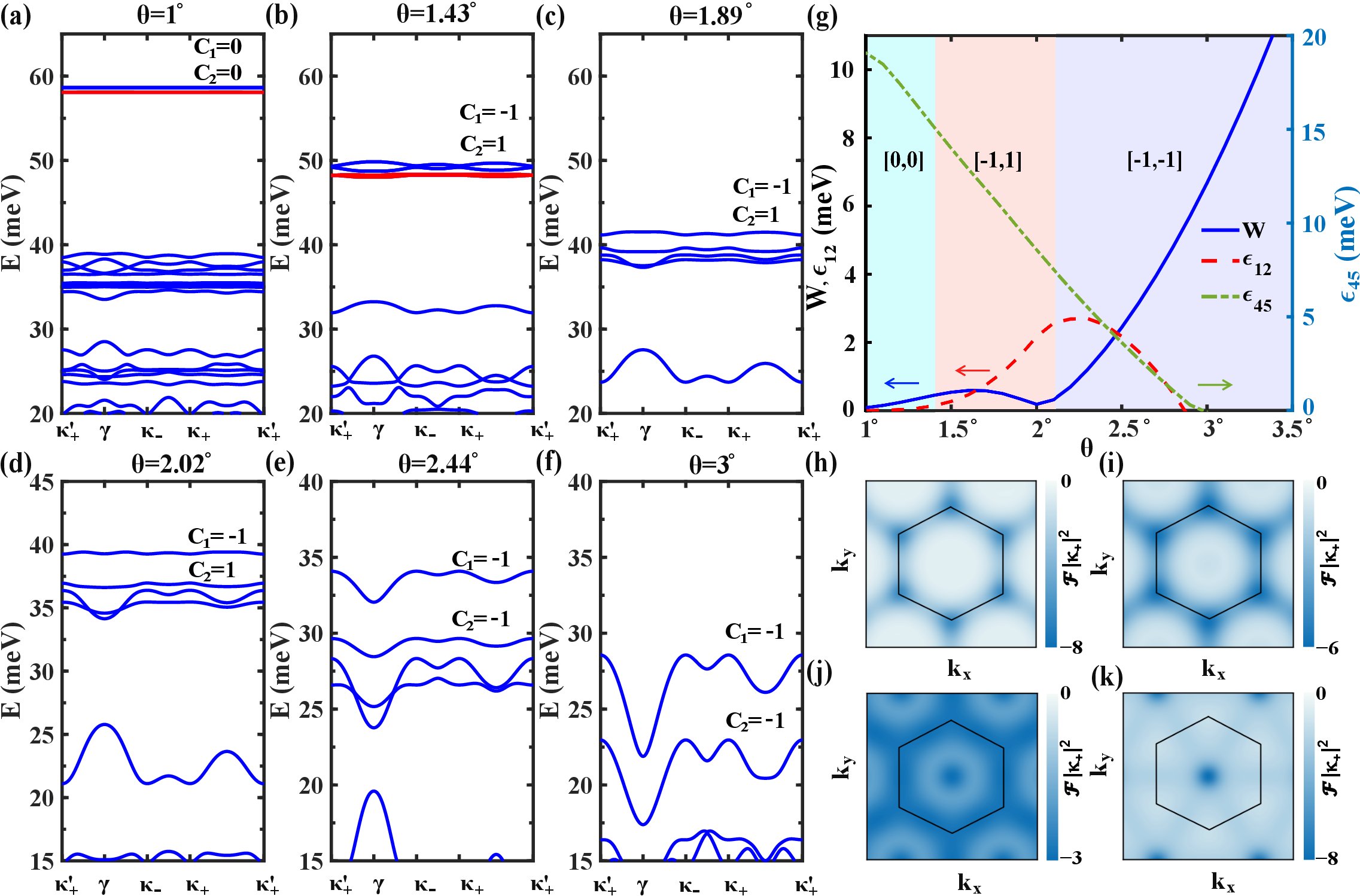}
		\caption{(a)-(f) The moir\'e band structures of AT6L-\textrm{MoTe$_2$} at different twist angles, corresponding to $\theta =1^{\circ},1.43^{\circ},1.89^{\circ},2.02^{\circ},2.44^{\circ},3^{\circ}$, respectively. In (a) and (b), the four topmost moir\'e flat bands can be approximately divided into two equivalent twisted TMD homobilayers (blue and red). (g) The bandwidth W of the 1st band, the gap $\epsilon_{12}$ between the 1st and 2nd bands, and the gap $\epsilon_{45}$ between the 4th and 5th bands as a function of $\theta$. (h)-(k) show the distribution of $\mathcal{F}$ for (h) $\theta =1.89^{\circ}$, (i) $2.02^{\circ}$, (j) $2.44^{\circ}$, and (k) $3^{\circ}$.
  }\label{6a} 
\end{figure*}

As well known, in order to realize FQAH state, the presence of a topological flat band is essential~\cite{Tang_prl236802_2011,Neupert_prl236804_2011,sheng_fractional_2011,Sun_PRL236803_2011,Regnault_PRX021014_2011,Justin_pnas_2015,Zhang_prb075127_2019,Crepel_PRB_2023,Morales_PRRL032022_2023}. And recent works indicates that the uniformity of Berry curvature may also play an important role~\cite{Devakul_magic_2021}.
Thus, we then discuss Berry curvature $\mathcal{F}(\mathbf{k})$ of the top flat band at different $\theta$. The $\mathcal{F}(\mathbf{k})$ in Fig.~\ref{4a}(h)-(k) corresponds to the band structures depicted in Fig.~\ref{4a}(c)-(f), respectively. With a small twist angle, e.g.~$\theta=1.89^\circ$ in Fig.~\ref{4a}(h), $\mathcal{F}$ is sharply peaked at the $\kappa_{\pm}$ points. As $\theta$ increases, $\mathcal{F}$ will first spread along the "bond" directions between the adjacent $\kappa_{\pm}$ points, as shown in Fig.~\ref{4a}(i) with $\theta=2.04^\circ$. Increasing $\theta$ further, $\mathcal{F}$ will shift from $\kappa_{\pm}$ to $\gamma$. Around $\theta=2.65^\circ$, $\mathcal{F}$ becomes almost uniform in the whole moir\'e Brillouin zone, see Fig.~\ref{4a}(j). This uniform distribution of $\mathcal{F}$ reminds us of Landau levels, which is also similar to the twisted $\mathrm{MoTe_2}$ homobilayers~\cite{Devakul_magic_2021}. In Fig.~\ref{4a}(k), we plot $\mathcal{F}$ at $\theta=3.48^\circ$, where $\mathcal{F}$ is peaked around $\gamma$ point. 

Based on the results above, we see that AT4L-\textrm{MoTe$_2$} shares similarities with the twisted $\mathrm{MoTe_2}$ homobilayers, while also exhibiting difference. The similarities are they both have a pair of topological flat bands, which behave similarly as $\theta$ increases. The differences are: (1) The AT4L-\textrm{MoTe$_2$} has a larger magic angle $\theta \approx 2.04^\circ$, while the magic angle of the twisted $\mathrm{MoTe_2}$ homobilayers is about $1.4^\circ$; (2) The Chern number of the two flat bands of the AT4L-\textrm{MoTe$_2$} are $[0,0]$ when $\theta<1.4^\circ$, but the Chern numbers of the twisted $\mathrm{MoTe_2}$ homobilayers are always nonzero.  

Most importantly, our results imply that AT4L-\textrm{MoTe$_2$} should also host FQAH state around the magic angle $\theta \approx 2.04^\circ$ like the twisted $\mathrm{MoTe_2}$ homobilayers, because that their moir\'e flat bands have nearly the same topological features, e.g.~the flat dispersion and uniformity of the Berry curvature near the magic angle.   
Meanwhile, owing to the similarity between AT4L-\textrm{MoTe$_2$} and twisted TMD homobilayers, this system could also serve as an ideal platform for realizing a Hubbard model~\cite{Topolo_Wu_prl2019,Devakul_magic_2021,Pan_Band_PRR2020,Pan_Quantum_PRB2020,ZhangYaHui_PRL2021,xu2022tunable}.

\subsubsection{AT6L-\textrm{MoTe$_2$}}
We then consider the case of AT6L-\textrm{MoTe$_2$}, where $N=6$. According to the decomposition rules in Eq.~\eqref{decouple},  its moir\'e bands near $E_f$ can approximately be viewed as originating from two independent twisted TMD homobilayers, i.e.~$h(\lambda_1)$ and $h(\lambda_2)$, with different scaling factors $\lambda_{1,2}$.  Here, each equivalent twisted TMD homobilayers will offer two moir\'e flat bands at $E_f$, thus, in this case, we should obtain four isolated moir\'e flat bands in the small twist angle region. 

The numerical results of  AT6L-\textrm{MoTe$_2$} are given in Fig.~\ref{6a}. In Fig.~\ref{6a} (a)-(f), we plot the moir\'e bands with different $\theta$. Clearly, as anticipated, the AT6L-\textrm{MoTe$_2$} does exhibit four moir\'e flat bands near $E_f$ at a small twist angle, which are separated from other bands by a large gap and nearly degenerate when $\theta<1.4^\circ$. As analyzed above, when $\theta$ is small [Fig.~\ref{6a} (a), (b)], the four moir\'e flat bands can be further approximately divided into two groups (blue and red), with two bands in each group, where each group here belongs to an equivalent twisted TMD homobilayers, i.e.~$h(\lambda_{1,2})$. For a large  $\theta$, this approximation does not work well, since the coupling between the two groups of moir\'e flat bands will obviously alter the band shape and the Chern number.  

Below, we only focus on the two topmost moir\'e flat bands out of the four. It is because, on one hand, the two topmost moir\'e flat bands are the most easily accessible bands in experiments, and on the other hand, these two bands can be isolated via adjusting $\theta$, allowing their Chern numbers to be well-defined. Thus, as illustrated in Fig.~\ref{6a} (g), we identify three distinct regions for $\theta$, according to the Chern numbers of the two topmost moir\'e flat bands:

When $\theta \lesssim 1.4^{\circ}$, the Chern numbers of the top two flat bands in this region are $[C_1,C_2]=[0,0]$. In this case, the four moir\'e flat bands are nearly degenerate near $E_f$, as depicted in Fig.~\ref{6a} (a), which are separated from other bands by a large gap. However, the first two moir\'e flat bands are actually isolated, albeit with very small gaps. Therefore, the calculations can yield a well-defined Chern number for the two topmost moir\'e bands here. In comparison with cases where $N<6$, the four nearly degenerate moir\'e flat bands in AT6L-\textrm{MoTe$_2$} will lead to a significantly larger DOS near $E_f$, potentially resulting in exotic correlated insulating states in this region.

When $1.4 \lesssim \theta \lesssim 2.1^{\circ}$, the top two flat bands exhibit non-zero  Chern numbers $[C_1,C_2]=[-1,1]$  in this range. First, as shown in  Fig.~\ref{6a} (b)-(d), the degeneracy of the moir\'e flat bands is lifted as $\theta$ increases. With a larger $\theta$, the gap between the two topmost moir\'e bands, namely $\epsilon_{12}$, becomes more apparent in this region, as illustrated in Fig.~\ref{6a} (g). Consequently, the 1st moir\'e flat band ($C_1=-1$) is isolated, as seen in Fig.~\ref{6a} (c) and (d). Meanwhile, the bandwidth of the 1st moir\'e band, $W$, remains small. In Fig.~\ref{6a} (g), we plot $W$ and $\epsilon_{12}$ as  functions of $\theta$. This indicates that $\theta \approx 2.02^\circ$ is a magic angle for the 1st moir\'e flat band, see Fig.~\ref{6a} (d), which is quite like the case of AT4L-\textrm{MoTe$_2$}. And the gap between the 4nd and 5rd moir\'e bands $\epsilon_{45}$ is also shown in Fig.~\ref{6a} (g) as well. As $\theta$ becomes larger, $\epsilon_{45}$ nearly monotonically decreases, and it becomes zero when $\theta$ approaches $3^\circ$.

When $\theta>2.1^\circ$, the Chern numbers of the top two moir\'e flat bands becomes $[C_1,C_2]=[-1,-1]$, as seen in Fig.~\ref{6a} (g). The cases with $\theta=2.44^\circ$ and $\theta=3^\circ$ are plotted in Fig.~\ref{6a} (e) and (f), respectively. In these cases, the top two moir\'e bands become more dispersive but remain isolated. 

The numerical results above imply that the top moir\'e flat band behaves very like those in twisted \textrm{MoTe$_2$} homobilayers and AT4L-\textrm{MoTe$_2$}, suggesting it is also an ideal platform for realizing FQAH states. We thus further study the distribution of the Berry curvature $\mathcal{F}$ of the top moir\'e flat band at different $\theta$, as illustrated in Fig.~\ref{6a} (h)-(k). At $\theta=1.89^\circ$, the peak of $\mathcal{F}$ distribution is located at the $\kappa_{\pm}$ points, see Fig.~\ref{6a} (h). 
When $\theta$ increases to the magic angle $2.02^\circ$, the $\mathcal{F}$ distribution will gradually shift along the edges of the moir\'e BZ, as illustrated in Fig.~\ref{6a} (i). When $\theta=2.44^\circ$, the distribution of $\mathcal{F}$ becomes uniform, see Fig.~\ref{6a} (j). At a large $\theta$, e.g.~$\theta=3^\circ$ in Fig.~\ref{6a} (k), $\mathcal{F}$ mainly distributes around the $\gamma$ points. The distribution of $F$ here is also very similar to that in twisted \textrm{MoTe$_2$} bilayer and AT4L-\textrm{MoTe$_2$}, suggesting that the moir\'e flat bands in these systems will share similar topological features, and thus may host FQAH states.  

\section{Summary}

In conclusion, we theoretically investigated the moir\'e band structures of ATML-\textrm{MoTe$_2$}, where the \textrm{MoTe$_2$} monolayers are stacked in an alternatingly twist way.  

We first propose an even-odd dependent decomposition rule for the ATML-\textrm{MoTe$_2$}, which works quite well when $\theta$ is small. The decomposition rules first indicate that an $N$-layer ATML-\textrm{MoTe$_2$} exhibits $N-2$ moir\'e flat bands near $E_f$, which relies on the important fact that the outermost two \textrm{MoTe$_2$} monolayers actually have little influence on the electronic structures near $E_f$, since their moir\'e potentials are significantly different from the middle ones. This decomposition rule also indicates that when $N$ is even (odd), ATML-\textrm{MoTe$_2$} can be approximately decoupled into several twisted TMD homobilayers (several twisted TMD homobilayers together with one twisted TMD heterobilayers). Meanwhile, we also reveal that the ATML-\textrm{MoTe$_2$} also has a mirror symmetry decomposition when $N$ is odd. 

Then, we numerically calculate the moir\'e band structures of the ATML-\textrm{MoTe$_2$} with $N=3,4,5,6$. All the numerical results are in good agreement with the proposed decomposition rules. Most importantly, our numerical calculations give three intriguing predictions:
 \begin{enumerate} 
 \item  The AT3L-\textrm{MoTe$_2$} ($N=3$) has one isolated moir\'e flat band, which corresponds to a triangular lattice Hubbard model, resembling the twisted TMD heterobilayers.  
 \item  The AT4L-\textrm{MoTe$_2$} ($N=4$) has two topological flat bands that are very similar to the twisted $\mathrm{MoTe_2}$ homobilayers, implying the possible existence of FQAH states.  
 \item  When $N>4$, the giant DOS induced by the multiple degenerate moir\'e flat bands may induce exotic correlated states.
\end{enumerate} 

Our works reveal that the ATML-\textrm{MoTe$_2$} is a promising moir\'e platform, which not only can be used to simulate the triangle lattice Hubbard model, but also may host the exotic FQAH states, analogous to those observed in twisted \textrm{MoTe$_2$} bilayer. We expect that our predictions can be immediately tested in future experiments.

\begin{acknowledgments}
We appreciate helpful discussions with Professor Fengcheng Wu. This work was supported by the National Natural Science Foundation of China (Grants No.~12141401), the National Key Research and Development Program of China (Grants No.~2022YFA1403501).
\end{acknowledgments}
		
		

\bibliography{refs}            

\end{document}